\documentclass[12pt]{article}

\usepackage{amssymb,amsmath,mathrsfs,epstopdf,slashed,color}
\usepackage{tikz}
\usetikzlibrary{matrix}
\usetikzlibrary{positioning}

\usepackage{graphicx}
\usepackage{placeins}
\usepackage{hyperref}
\usepackage{cite}

\allowdisplaybreaks[1]

\makeatletter

\@addtoreset{equation}{section}
\makeatother

\newcommand{\vol}{\mathrm{vol}}

\title{\bf Why the Cosmological Constant is so Small ? A String Theory Perspective}
\author{S.-H. Henry Tye}

\begin{document}

\begin{titlepage}

\setcounter{page}{0}
  
\begin{flushright}
 \small
 \normalsize
\end{flushright}

\vskip 1.6cm
\begin{center}

{\Large \bf Why the Cosmological Constant is So Small ? \\
A String Theory Perspective}  

\vskip 1.8cm
  
{\large S.-H. Henry Tye}
 
 \vskip 0.6cm

 Jockey Club Institute for Advanced Study and Department of Physics, \\
 Hong Kong University of Science and Technology, Hong Kong\\
 Laboratory for Elementary-Particle Physics, Cornell University, Ithaca, NY 14853, USA

 \vskip 1cm

{\bf ~~~~with Yoske Sumitomo and Sam Wong}

\vskip 1cm
  
\abstract{\normalsize
With no free parameter (except the string scale $M_S$), dynamical flux compactification in Type IIB string theory determines both the cosmological constant (vacuum energy density) $\Lambda$ and the Planck mass $M_P$ in terms of $M_S$, thus yielding their relation. Following elementary probability theory, we find that a good fraction of the meta-stable de Sitter vacua in the cosmic string theory landscape tend to have an exponentially small cosmological constant $\Lambda$ compared to either the string scale $M_S$ or the Planck scale $M_P$, i.e., $\Lambda \ll M_S^4 \ll M_P^4$.  Here we illustrate the basic stringy ideas with a simple scalar field $\phi^3$  (or $\phi^4$) model coupled with fluxes to show how this may happen and how the usual radiative instability problem is  bypassed (since there are no parameters to be fine-tuned). These low lying semi-classical de Sitter vacua tend to be accompanied by light scalar bosons/axions, so the Higgs boson mass hierarchy problem may be ameliorated as well.
}

\vspace{1cm}

{\it ~~~~~~~~~~~~~~~~~~~~~~~{\bf Talk given at String 2016, Beijing (updated)}}

\vspace{0.4cm}

{\it ~~~~~~~ to appear in {\bf Advances in Theoretical and Mathematical Physics}}

\vspace{0.7cm}
\begin{flushleft}
 Email: \href{mailto: iastye@ust.hk}{iastye at ust.hk}
\end{flushleft}
 
\end{center}
\end{titlepage}

\setcounter{page}{1}
\setcounter{footnote}{0}


\parskip=5pt

\section{Introduction}

Cosmological data strongly indicates that our universe has a vanishingly small positive cosmological constant $\Lambda$ (or vacuum energy density) as the dark energy,
\begin{equation}
\Lambda \sim 10^{-122}M_P^4
 \label{L1}
\end{equation}
where the Planck mass  $M_P = G_N^{-1/2}\simeq 10^{19}$ GeV.  Such a small $\Lambda$ is a major puzzle in physics.\footnote{ if the dark energy is due to some other mechanism, e.g., quintessence, then the cosmological constant may have to be even smaller, or a more fine-tuned cancellation/correlation must be present.}
In general relativity, $\Lambda$ is a free arbitrary parameter one can introduce, so its smallness can be accommodated but not explained within quantum field theory. 
On the other hand, string theory has only a single parameter, namely the string scale $M_S=1/\sqrt{2 \pi \alpha' }$, so everything else should be calculable for each string theory solution. String theory has 9 spatial dimensions, 6 of them must be dynamically compactified to describe our universe. Since both $M_P$ and $\Lambda$ are calculable, $\Lambda$ can be determined in terms of $M_P$ dynamically in each local minimum compactification solution. This offers the possibility that we may find an explanation for a very small positive $\Lambda$. This happens if a good fraction of the meta-stable deSitter (dS) vacua in the landscape tend to have a very small $\Lambda$, as is the case in the recent studies in flux compactification in string theory. Here, instead of reporting the main result of our work in the past few years \cite{Sumitomo:2012wa,Sumitomo:2012vx,Sumitomo:2012cf,Sumitomo:2013vla,Tye:2016jzi}, I like to introduce the basic stringy idea in a simple $\phi^3$ or $\phi^4$ quantum field theory model for illustration. 

Although string theory has no free parameter (i.e., masses and couplings), it has fluxes and moduli in flux compactification from 10 spacetime dimensions to 4 (or any lower than 10) spacetime dimensions. In low energies, the moduli (plus the dilaton) are scalar fields describing the shape and size of the compactified manifold. Fluxes come from  anti-symmetric P-form field strengths $F^{\mu_1 \mu_2 . . . . \mu_P}$  where $\mu_0, \mu_1, \mu_2, \mu_3$ are the 4-dimensional spacetime indices and the rest stand for the internal dimensions. (For $P < 4$, consider its dual.) So the effective 4-dimensional $F_i=F^{0123}_i, \quad  i=1,2,...,N$, refers to the collection of internal indices. Each $F_i$ takes a constant value in 4-dimensional spacetime. They contribute to the vacuum energy density. 

Bousso and Polchinski show that each $F_i$ takes only quantized values at the local minima in string theory \cite{Bousso:2000xa}. For enough number of such fluxes (say $N > 14$), with each $F_i$ sweeping through a range of discrete values, the spacing between allowed vacuum energy density values is small enough so a small $\Lambda$ like that observed (\ref{L1}) can be one of the allowed values. They call this "dense discretuum". Our approach is to include the moduli coupled to the fluxes and find the vacuum energy density $\Lambda$ at every local minimum. Sweeping over all choices of the flux values, we then find the distribution of the $\Lambda$ values. Here, dynamics is brought in beyond a simple counting exercise, since not all choices of fluxes have a local minimum, while some flux choices yield multiple local minima.  

Start with the four-dimensional low energy (supergravity) effective potential $V (F_i, \phi_j)$ obtained from flux compactification in string theory, where  $F_i$ are the 4-form field strengths and $\phi_j$ are the complex moduli (and dilaton) describing the size and shape of the compactified manifold as well as the couplings.
In the search of classical minima, this flux quantization property allows us to rewrite $V( F_i, \phi_j)$ as a function of  the quantized values $q_in_i$ of the fluxes present, 
$$V( F_i, \phi_j) \rightarrow V(q_in_i, \phi_j), \quad \quad i=1,2,...,N, \quad j=1,2,..., K$$
where the charges $q_i$ are determined by the compactification dynamics. (To simplify the discussion, we shall suppress the $q_i$ so $n_i$ takes discrete instead of integer values.)
Since string theory has no continuous free parameter, there is no arbitrary free parameter in $V(n_i, \phi_j)$, though it does contain (in principle) calculable quantities like $\alpha'$ corrections, loop and non-perturbative corrections, and geometric quantities like Euler index $\chi$ etc.. 

For a given set of discrete flux parameters $\{n_i\}$, we can solve $V(n_i, \phi_j)$ for its meta-stable (classically stable) vacuum solutions via finding the values $\phi_{j, {\rm min}} (n_i)$ at each solution and determine its vacuum energy density $\Lambda=\Lambda(n_i, \phi_{j, {\rm min}} (n_i))=\Lambda (n_i)$.  Since we are considering the physical $\phi_j$, it is the physical $\Lambda$ we are determining.  Since a typical flux parameter $n_i$ can take a large range of discrete values, we may simply treat each $n_i$ as an independent random variable with some  distribution $P_i(n_i)$. Collecting all such solutions, we can next find the probability distribution $P(\Lambda)$ of $\Lambda$ of these meta-stable solutions as we sweep through all the discrete flux values $n_i$.  That is, putting $P_i(n_i)$ and $\Lambda (n_i)$ together yields $P(\Lambda)$, 
\begin{equation}\label{pf}
P(\Lambda)= \sum_{n_i} \delta(\Lambda - \Lambda (n_i))  \Pi_i P_i(n_i) 
\end{equation}
where $\sum_{n_i} P_i(n_i) =1$ for each $i$ implies that $\int P(\Lambda) d \Lambda =1$.  For large enough ranges for $n_i$, we may treat each $P_i(n_i)$ as a smooth continuous function over an appropriate range of values. 

Simple probability properties show that $P(\Lambda)$ easily peaks and diverges at $\Lambda=0$ \cite{Sumitomo:2012wa}, implying that a small $\Lambda$ is statistically preferred. For an exponentially small $\Lambda$, the statistical preference for $\Lambda \simeq 0$ has to be overwhelmingly strong, that is, the properly normalized $P(\Lambda)$ has to diverge (i.e., peak) sharply at $\Lambda =0$. Such an analysis has been applied to the K\"ahler uplift scenario in Type IIB string theory \cite{Rummel:2011cd}, where $P(\Lambda)$ is so peaked at $\Lambda=0$ that the the median $\Lambda$ matches the observed $\Lambda$ (\ref{L1}) if the number of complex structure moduli $h^{2,1} \sim{ \cal O} (100)$ \cite{Sumitomo:2012vx}.  Such a value for $h^{2,1}$ is quite reasonable for a typical manifold considered in string theory. A study of the Racetrack K\"ahler uplift model also yields an exponentially small median for $\Lambda$ \cite{Sumitomo:2013vla}. That is, an overwhelmingly large number of meta-stable vacua have an exponentially small $\Lambda$, and hardly any vacuum has a value close to the string/Planck scale. So statistically, we should end up in a vacuum with an exponentially small $\Lambda$. That is, a very small $\Lambda$ is quite natural. 

In usual quantum field theory, the parameters of any model (masses, couplings and $\Lambda$) that include the standard model of strong and electroweak interactions have to be fine-tuned to satisfy Eq.(\ref{L1}). However, quantum corrections are typically orders-of-magnitude bigger than the observed value. So the renormalized parameters have to be re-fine-tuned after each order of radiative correction. This is the radiative instability problem.
Since there are no free parameters to be fine-tuned in string theory, the radiative instability problem is simply absent (or bypassed) here.

This leads us to conjecture that

 {\it Most meta-stable vacua in regions of the cosmic stringy landscape have $|\Lambda| \ll M_P^4$. }

If true, this may (1) provide an explanation why the observed $\Lambda$ is so small, and (2) after inflation, why the universe is not trapped in a relatively high $\Lambda$ vacuum. 
Before a very brief review of the string theory models, let us discuss in some detail an illustrative $\phi^3/\phi^4$ model, where there is no uncoupled sector and all couplings/parameters are treated as if they are flux parameters so they will take random values within some reasonable ranges \cite{Sumitomo:2012vx,Tye:2016jzi}. There, the physical (loop corrected) $V(n_i, \phi_j)$ yields $P(\Lambda_{\rm ph})$ for the physical $\Lambda_{\rm ph}$ while the tree (or bare) $V(n_i, \phi_j)$ yields $P(\Lambda_0)$ for the tree $\Lambda_0$. We find that $P(\Lambda_{\rm ph})$ hardly differs from $P(\Lambda_0)$. Both $P(\Lambda)$s peak (i.e., diverge) at $\Lambda=0$, and the two sets of statistical preferred flux values for $\Lambda \sim 0$ are in general only slightly different. In fact, up to two-loops, $P(\Lambda_{\rm ph})$ is essentially identical to the tree $P(\Lambda_0)$. As a result, although radiative instability may be present for any fixed flux choice, the statistical preference approach actually evades or bypasses this radiative instability problem. We like to convince the readers that this phenomenon of bypassing the radiative instability problem stays true in more complicated models, and may well be applied to very light scalar boson masses (if present). 
 
There are a vast number of very small $\Lambda$ dS vacua in the cosmic string landscape. Lest one may think the accumulation of $\Lambda \simeq 0^+$ is due to energetics (i.e., small positive $\Lambda$s are energetically preferred over not so small positive $\Lambda$s), we note that the same accumulation happens for AdS vacua as well; that is, $P(\Lambda)$ peaks (diverges) as $\Lambda \rightarrow 0^-$ \cite{Sumitomo:2012cf}. Our universe rolling down the landscape after inflation is unlikely to be trapped by a relatively high dS vacuum, since there is hardly any around. However, since it has to pass through the positive $\Lambda$ region first, it is likely to be trapped at a small positive $\Lambda$ vacuum (as there are many of them) before reaching the sea of AdS vacua with small negative $\Lambda$. This scenario also implies that the vacuum we are living in today is only meta-stable. Fortunately, simple estimate indicates that its lifetime can easily be much longer than the age of the universe.

\section{An Illustrative $\phi^3/\phi^4$ Toy Model}

The statistical preference for a small $\Lambda$ typically follows if the low energy effective potential has (1) no continuous free parameter and (2) all sectors are connected via interactions, as is the case in string theory; that is, it is a function of only scalar fields or moduli, quantized flux values, discrete values like topological indices, and calculable quantities like loop and string corrections, with no disconnected sectors.  To get some feeling on some of these features, let us review the single scalar field $\phi$ polynomial model. In this model, gravity and so $M_P$ is absent. So the statistical preference for a small $\Lambda$ shows up only as the (properly normalized) probability distribution $P(\Lambda)$ peaks at $\Lambda=0$, in particular when $P(\Lambda)$ diverges there, i.e., 
  \begin{equation}\label{phipeak}
\lim_{\Lambda \rightarrow 0^+} P(\Lambda) \rightarrow \infty 
 \end{equation}
The divergence of $P(\Lambda=0)$ in this toy model is rather mild here, so it is far from enough to explain the very small observed value of $\Lambda$ (\ref{L1}); but it does allow us to explain a few properties that are relevant for later discussions.
Consider the tree level potential,
  \begin{equation}
  \label{A11}
V_0(\phi) = a_1 \phi + {a_2 \over 2} \phi^2 + {a_3 \over 3!} \phi^3 +\frac{a_4}{4!}\phi^4
 \end{equation}
 where $\phi$ is a real scalar field, mimicking a modulus. 
     
Imposing the constraint that the tree level  $V_0$ has no continuous free parameter except some scale $M_s$, the parameters $a_1$, $a_2$, $a_3$ and $a_4$ mimic the flux parameters that take only discrete values of order of the $M_s$ scale, thus spanning a ``mini-landscape". Let them take only real values for simplicity. We may also choose units so $M_s=1$. For a dense enough discretuum for each flux parameter, a flux parameter may be treated as a random variable with continuous value over some range. Let us look for dS solutions with flux parameters
$a_1$, $a_2$, $a_3$, $a_4 \in [-1, 1]$
or some other reasonable ranges. We start with the tree-level properties and then discuss the multi-loop corrections. 

\begin{figure}
\begin{center}
\includegraphics[scale=0.8]{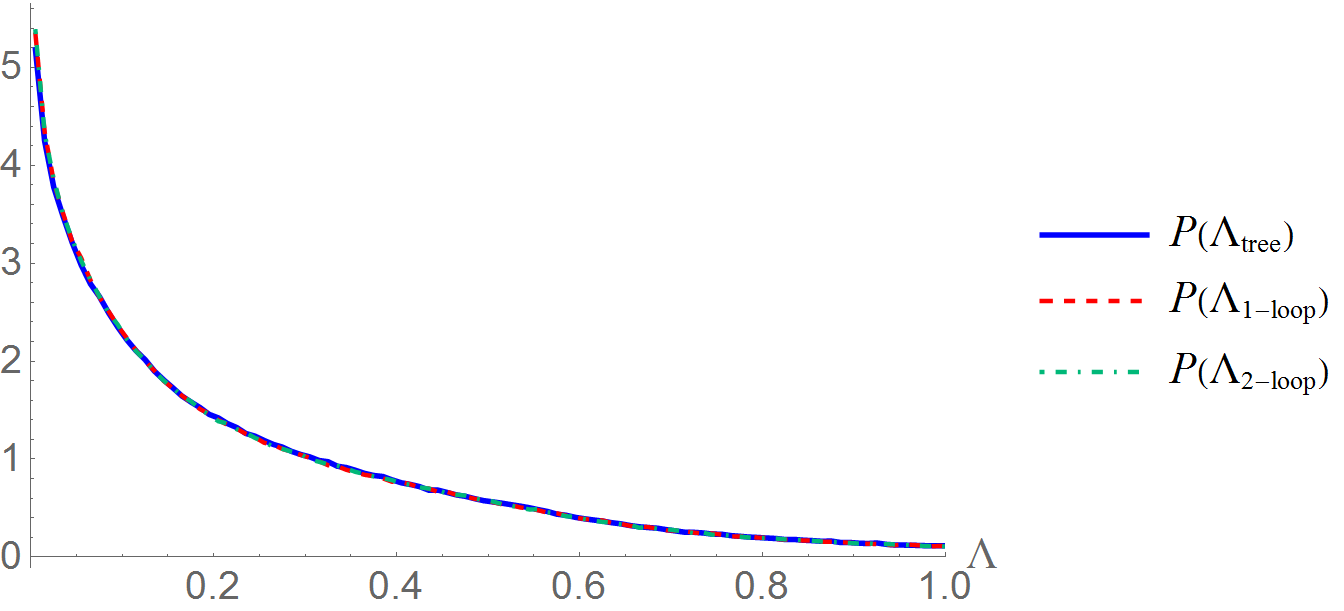}
\end{center}
\caption{Probability distribution $P(\Lambda)$ for the $\phi^4$ model (similar behaviors for the $\phi^3$ model) for the tree-level and the one- and two-loop corrected cases. The blue solid curve is for the tree-level $P(\Lambda_0)$, the red dashed curve is for the one-loop corrected $P(\Lambda_1)$ and the green dot-dash curve is for the two-loop corrected $P(\Lambda_2)$. We see that the loop-corrected and the tree $P(\Lambda)$s are essentially on top of each other, showing that loop corrections have little impact on the distribution $P(\Lambda)$. In particular, the peaking behavior of $P(\Lambda)$ at $\Lambda=0$ remains intact. 
} \label{randomc}
\end{figure}

Starting with the tree-level effective potential $V_0(\phi)$ (\ref{A11}), we impose the stability $M^2=\partial_\phi^2 V_0|_{v_0} >0$ at the extremal points given by $\partial_\phi V_0{\big |}_{v_0}=0$, with each vacuum expectation value $v_0$ yielding $\Lambda_0(v_0) = V_0(v_0)$ and
  \begin{equation}
  \label{A22}
  M_0^2=\frac{\partial^2 V_0}{\partial \phi^2}{\big |}_{v_0} = +a_2+a_3v_0 + a_4v_0^2/2 >0
  \end{equation}
  with $\lambda =\frac{\partial M^2}{\partial v}=a_3+a_4v_0$.
  Not all choices of fluxes yield a local minimum, while some choices yield more than one minimum.
We study three case : the $\phi^3$ model with $a_3=1$ and with random $a_3$, and the $\phi^4$ case with random flux parameters $\{a_i\}$. Using Eq.(\ref{pf}), we find $P(\Lambda)$. The $\phi^4$ case is shown in Figure \ref{randomc}. $P(\Lambda)$ clearly peaks (i.e, diverges logarithmically) at $\Lambda=0$. (Similar divergence of $P(\Lambda)$ emerges for the $\phi^3$ case.)

A few comments are in order here :

$\bullet$ $V_0(\phi)$ (\ref{A11}) has no free parameter, as demanded by string theory. Here we have chosen a flat distribution for each flux parameter. We do expect the distributions $P_i(a_i)$ to be smooth and include $a_i=0$, since zero flux value must be permitted. Actually, the peaking of $P(\Lambda)$ is independent of the precise distributions of $P_i(a_i)$, as long as they are relatively smooth.  

$\bullet$ We are not allowed to introduce a ``constant" or an independent flux parameter $a_0$ term by itself in $V_0(\phi)$ (\ref{A11}), since it will be disconnected to the $\phi$ terms in $V_0(\phi)$. In string theory, everything is coupled to everything else, directly or indirectly, at least via the closed string sector, which includes gravity. For example, the following possibility is dis-allowed (i.e., in the swampland),
$$V(\phi)= V_1(\phi_1) + V_2(\phi_2)$$
if $V_1$ and $V_2$ are totally disconnected. In this case, we can minimize $V_2(\phi_2)$ (for fixed flux values in $V_2(\phi_2)$) to obtain  $\Lambda_2$ so
$$V(\phi)= V(\phi_1) + \Lambda_2$$
where $\Lambda_2$ appears as a constant term in $V(\phi)$ un-connected to $\phi_1$ or the flux parameters in $V_1(\phi_1)$. In this case (dis-allowed by string theory), even if both $P_1(\Lambda_1)$ and $P_2(\Lambda_2)$ peak at their respective zeros, $P(\Lambda)$ does not peak at $\Lambda=\Lambda_1+\Lambda_2=0$.
 
$\bullet$ It is easy to show that $P(\Lambda)$ similarly peaks at $\Lambda=0$ if we include higher terms (e.g., up to a $a_6\phi^6/6!$ term) in $V_0(\phi)$ (\ref{A11}).

$\bullet$ For vacuum energy, we may choose to introduce the kinetic $F_i^2/2$ terms to $V_0(\phi)$ (\ref{A11}); for example,
$$E= V(\phi) + \frac{1}{2} a_1^2$$
in the $\phi^3$ case, where the flux term takes constant discrete values. We see that $P(\Lambda)$ again peaks at $\Lambda=0$. Notice that such a flux term is coupled to $\phi$ in $V(\phi)$. In actual string theory models, $F_i = q_in_i$ where, although $n_i$ is an integer,  the "charge $q_i$" is determined by dynamics, probably some function of the moduli. So a pure $F_i^2$ implies that some (presumably heavier)  moduli have already been stabilized to yield the charge $q_i$. Adding more "kinetic" flux terms into $E$ without additional couplings and/or moduli (scalar fields) may erase the peaking property of $P(\Lambda)$.

$\bullet$ Here we find the accumulation of vacua with $\Lambda \simeq 0^+$. We note that $P(\Lambda)$ also peaks (diverges) as $\Lambda \rightarrow 0^-$ in this toy model. 
In fact, typical AdS solutions of $V(n_i, \phi_j)$ in supergravity models involve 2 branches: supersymmetric vacua and non-supersymmetric vacua, where the latter set mirrors the dS solutions (see e.g., \cite{Sumitomo:2012cf}). So, for a given range of small $|\Lambda|$, we expect more AdS vacua than dS vacua; that happens even before we relax the constraint to allow light tachyons which do not destabilize the AdS vacua. 

$\bullet$ Let us discuss the multi-loop contributions to the effective potential $V(\phi)$. We see that the $n$-th loop contribution to $V(\phi)$, namely $V_n(\phi)$, is a function of $M^2(\phi)$, $\lambda(\phi)$ (\ref{A22}), and $a_4$ only. At the one-loop level, $V_1(\phi)$ is a function of $M^2(\phi)$ only. 
Simple dimensional reasoning yields
  \begin{align}
 V(\phi) = V_0+ \sum_{n \ge1} V_n = M^4F\left(\frac{\lambda^2}{M^2}, \ln ({M^2}), a_4\right),\nonumber \\ M^2(\phi)=V''_0(\phi),\quad \lambda(\phi)=V'''_0(\phi), \quad a_4(\phi)=V''''_0(\phi),
   \end{align}
where each prime stands for a derivative with respect to $\phi$ and $F(\lambda^2/M^2, \ln(M^2), a_4)$ is a polynomial in the dimensionless parameters $\lambda^2/M^2, \ln(M^2)$ and $a_4$. More precisely, for $n \ge 1$,
$$V_n = \frac{M^4}{(4 \pi)^{2n}} f_n \left({\lambda^2}/{M^2}, \ln ({M^2}), a_4\right) $$
where $f_n$ is a polynomial up to $n$-th power in $\ln (M^2)$, and $(n-1)$-th (combined) power in ${\lambda^2}/{M^2}$ and $a_4$, with $n$-dependent coefficients which grow much slower than the $(4 \pi)^{2n}$ factor. 

The key of a naturally small $\Lambda_{\rm ph}$ depends on its functional dependence on the flux values, which is different from that for $\Lambda_0$. Here we consider explicitly the one- and two-loop contributions to $\Lambda$ and find the $P(\Lambda_{\rm ph})$ for the one- and two-loop corrected cases, namely $P(\Lambda_1)$ and $P(\Lambda_2)$. At least up to two-loops, $P(\Lambda_{\rm ph})$ continues to peak (diverge) at  $\Lambda_{\rm ph}=0$. In fact, the loop corrected $P(\Lambda)$ are essentially indistinguishable from the tree $P(\Lambda_{0})$, as shown in FIG \ref{randomc}. This is despite the fact that, for a specific choice of $\{a_i\}$ that yields $\Lambda_0=0$, $\Lambda_1 \ne 0$ and $\Lambda_2 \ne 0$ in general. Similarly, for a specific choice of $\{a_i\}$ that yields $\Lambda_1=0$, $\Lambda_0 \ne 0$ and $\Lambda_2 \ne 0$ in general. This radiative instability issue is absent in the string landscape since we sweep through all choices of $\{a_i\}$.

To summarize, the statistical preference for $\Lambda=0$ is robust, for either the tree-level $\Lambda_0$ or the loop-corrected $\Lambda_{\rm ph}$. Although the functional dependence of $\Lambda$ on the flux parameters are different for $\Lambda_0$ and $\Lambda_{\rm ph}$,  
nevertheless, given the same probability distributions for the flux parameters, we see that $P(\Lambda_{\rm ph})$ is essentially the same as  $P(\Lambda_0)$.
It will be nice to investigate the above properties for more general quantum field theory models that satisfy the stringy conditions : no free parameters except flux parameters and no uncoupled sectors. Of course, the cases we are really interested in are the flux compactifications in string theory. However, we do gain some intuitive understanding in examining this simple model. 

Actually we are interested only in the preferred values of the physical $\Lambda$. However, including quantum effects fully is in general a very challenging problem in any theory. Fortunately, if one can argue that the peaking behavior of $P(\Lambda)$ is hardly modified by quantum corrections, as this model suggests, a simpler tree-level result provides valuable information on the statistical preference of a small physical $\Lambda$. For ground states in string theory, an effective potential description may be sufficient to capture the physics of the value of $\Lambda$ in some regions of the landscape.  We may hope that stringy corrections will not qualitatively disrupt the statistical preference approach adopted here.

\subsection{Scalar Masses in the Toy Model}
 
 We also like to propose that 
 
{\it A dS vacuum with a naturally small $\Lambda$ tends to be accompanied by light scalar bosons.}
 
 Consider the 4-dimensional effective action
\begin{equation}
  \label{Action0}
S=\int dx^4 \sqrt{-g}\left[-\Lambda +\frac{M_P^2}{16 \pi}R - \frac{m_H^2}{2}\Phi_H^2 +  . . .\right]
 \end{equation}
where we have displayed all the relevant operators that are known to be present in nature. If we ignore the $\Lambda$ (the most relevant operator) term, then we have two scales, $M_P \gg m_H$. Why the Higgs mass $m_H$ is so much smaller than the Planck mass $M_P$ poses the well-known mass hierarchy problem. Now knowing that a very small $\Lambda$ is present in nature, we like to know its origin. If its value arises via fine-tuning or by pure accident, we have to consider $M_P$ as more fundamental and so are led back to the original mass hierarchy problem.  However, if the smallness of $\Lambda$ arises naturally, in that most of the de Sitter vacua in string theory tend to have a very small $\Lambda$, we should expect some scalar masses comparable to the $\Lambda$ scale, as is the case in the string models examined.  Following this viewpoint, we may instead wonder why the Higgs mass is so much bigger than $\Lambda$, i.e., $m^2_H \gg \Lambda/M_P^2$ (an inverted mass hierarchy problem). Surely, we should re-examine the mass hierarchy problem in this new light. After all, $\Lambda$ is the most relevant operator here.

Along this direction, we show that the following scenario can easily happen : the physical mass-squared probability distribution $P_j(m_j^2)$ for some scalar field $\phi_j$ may be peaked at $m_j^2=0$ but the peaking is less strong than that for
 $\Lambda$. If the Higgs boson is such a particle, i.e., $\Phi_H=\phi_j$, then it is natural for 
  \begin{equation}
 \Lambda/M_P^2 \ll m_H^2 \ll M_S^2 \ll M_P^2 
 \label{HiggsSPA}
  \end{equation}
This statistical preference approach allows us to circumvent the original mass hierarchy problem; that is, a small Higgs mass is natural, not just technically natural. 

It is straightforward to find that $P(m^2)$ does not peak at $m^2=0$ in the above $\phi^3/\phi^4$ model, in every case considered, loop corrected or not. 
In more realistic models, $P(\Lambda_{\rm ph})$ has to diverge at $\Lambda_{\rm ph}=0$ much more sharply than the logarithmical divergence shown in this model.
In the non-trivial models in string theory studied so far, we see that both $P(\Lambda_{\rm ph})$ for $\Lambda_{\rm ph}$ and $P(m^2_{\rm ph})$ for the some bosons prefer small values, while the peaking in $P(\Lambda_{\rm ph})$ is much stronger than that in $P(m^2_{\rm ph})$.  
If one applies this to the Higgs boson in a phenomenological model, the observed situation (\ref{HiggsSPA}) can follow from their statistical preferences.  

The way of bypassing the radiative instability problem should also apply to the masses as well when the probability distribution $P(m^2)$ for some scalar mass also peaks at $m^2=0$. Furthermore, one may convince oneself that this statistical preference for a small $\Lambda$ also bypasses the disruptions caused by phase transitions during the evolution of the early universe, as the universe rolls down the landscape after inflation in search of a meta-stable minimum. (See Discussion below.)

\section{A K\"ahler Uplift Model of Flux Compactification}

Here we summarize the results of a flux compactification model where the AdS vacua are K\"ahler uplifted to dS vacua via the presence of an $\alpha'^3$ correction plus a non-perturbative term. Using reasonable probability distributions for the flux values, it has been shown that the probability distribution $P(\Lambda)$ peaks sharply at $\Lambda=0$, resulting in a median $\Lambda$ comparable to the observed value if the number of complex structure moduli $h^{2,1} \sim {\cal O}(100)$.

To be specific, consider a Calabi-Yau-like three-fold $M$ with a single ($h^{1,1}=1$) K\"ahler modulus and a relatively large $h^{2,1}$ number of complex structure moduli, so the manifold $M$ has Euler number $\chi(M)=2(h^{1,1}-h^{2,1}) <0$.
The simplified model of interest was first studied in Ref\cite{Rummel:2011cd}, which incorporates a number of relevant features known in the literature and where earlier references can be found. Setting $M_P=1$,
\begin{equation}
 \begin{split}
  V =& e^{K} \left(K^{I \bar{J}} D_I W D_{\bar{J}} {\overline W} - 3\left|W \right|^2\right),\\
  K =& K_{\rm K} + K_{\rm d} + K_{\rm cs}= -2    \ln \left({\cal V} + {\hat{\xi} \over 2} \right) -    \ln \left(S+\bar{S} \right) -  \sum_{i=1}^{h^{2,1}} \ln \left(U_i + \bar{U}_i  \right),\\
  {\cal V} \equiv& {\vol \over \alpha'^3 } =  (T + \bar{T})^{3/2},  \quad 
  \hat{\xi} =  -\frac{\zeta(3)}{4\sqrt{2}(2\pi)^3} \chi(M) \left( S + \bar{S} \right)^{3/2}>0, \\
  W =&  W_0(U_i,S) +  A e^{-a T}, \\
   W_0(U_i,S) =&  c_1 +\sum_{i=1}^{h^{2,1}} b_i U_i - S \left(c_2 + \sum_{i=1}^{h^{2,1}} d_i U_i\right) 
   +\sum_{i,j}^{h^{2,1}} \alpha_{ij}U_iU_j,
 \end{split}
 \label{LVS}
\end{equation}
The original model has $h^{2,1}=3$ complex structure moduli and $\alpha_{ij}=0$. We simply take this form and straight-forwardly generalize it to an arbitrary number of complex structure moduli. In known models for  $h^{2,1}>3$, the potential is actually somewhat different. However, this straightforward generalization, though naive, allows us to solve the model semi-analytically, which is crucial to obtain any numerical properties when $h^{2,1}$ is large \cite{Sumitomo:2013vla}. In this sense, our model is at best semi-realistic.

The flux contribution to $W_0 (U_i,S)$ depends on the dilation $S$ and the $h^{2,1}$ complex structure moduli $U_i$ ($i=1,2,..., h^{2,1}$), while the non-perturbative term for the K\"ahler modulus $T$ is introduced in the superpotential $W$. The dependence of $A$ on $U_i, S$ are suppressed.
The model also includes the $\alpha'$-correction (the $\hat{\xi}$ term) to the K\"ahler potential where $c_i, b_i$,  $d_i$ and the non-geometric $\alpha_{ij}=\alpha_{ji}$ are (real) flux parameters that may be treated as independent random variables with smooth probability distributions that allow the zero values. Here we are interested in the physical $\Lambda$ (instead of, say, the bare $\Lambda$), so the model should include all appropriate non-perturbative effects, $\alpha'$ corrections as well as radiative corrections.
We see that the above simplified model (\ref{LVS}) includes a non-perturbative $A$ term to stabilize the K\"ahler modulus and the $\alpha'$ correction $\hat \xi$ term to lift the solution to de-Sitter space. In the same spirit, all  parameters in the model, in particular the coupling parameters $c_i, b_i$, $d_i$ and $\alpha_{ij}$ in $W_0$ (\ref{LVS}), should be treated as physical parameters that have included all relevant corrections.

Now we sweep through the flux values $c_i, b_i$ and $d_i$ treating them as independent random variables to find the probability distribution $P(\Lambda)$. The ranges of flux values are constrained by our weak coupling approximation (i.e., $s>1$) et. al.. For any reasonable probability distributions $P_i(c_i)$, $P_i(b_i)$ and $P_i(d_i)$, we find that $P(\Lambda)$ peaks (and diverges) at $\Lambda=0$. To quantify this peaking behavior,  it is convenient to summarize the result by looking at $\Lambda_{Y \%}$.  That is, there is $Y \%$ probability that $\Lambda_{Y \%} \ge \Lambda  \ge 0$. So $\Lambda_{50 \%}$ is simply the median. 
In Ref.\cite{Sumitomo:2012vx}, we find that (with $\alpha_{ij}=0$), as a function of the number $h^{2,1}$ of complex structure moduli, for $h^{2,1} > 5$ and $\Lambda \ge 0$,
\begin{align}
\label{lambdam}
\Lambda_{50 \%} &\simeq 10^{-h^{2,1} -2} M_P^4 \nonumber  \\
\Lambda_{10 \%} &\simeq 10^{-1.3 h^{2,1} -3} M_P^4 \nonumber\\
\langle\Lambda \rangle& \simeq 10^{-0.03 h^{2,1} -6} M_P^4
\end{align}
where we have also given $\Lambda_{10 \%}$. We see that the average $\langle\Lambda \rangle$ does not drop much, since a few relatively large $\Lambda$s dominate the average value.
A typical flux compactification can have dozens or even hundreds of $h^{2,1}$, so we see that a $\Lambda$ as small as that observed in nature can be dynamically preferred. 
Scaler masses are of order \cite{Tye:2016jzi},
$$m^2M_P^2 /\Lambda \sim {\cal O}(1)$$
while axions are massless before non-perturbative interactions are included. Turning on $\alpha_{ij}$ tend to lift the moduli masses somewhat. Presumably there are other heavier scalar bosons that have been integrated out in this model. Although their masses are by definition much heavier than the moduli masses just discussed, some of them  can easily have masses that are much smaller than the string scale $M_S$.

\section{Modulus Masses in Racetrack K\"ahler Uplift}

The K\"ahler uplift model studied in the last section has a single non-perturbative term in the superpotential $W$. To relax the constraint on the volume size, we generalize the model to include two non-perturbative terms in $W$, i.e., the well-studied {\it racetrack} model. (See Ref.\cite{Sumitomo:2013vla} for details.)
Unlike the K\"ahler uplift model studied previously, the $\alpha'$-correction is more controllable for the meta-stable de-Sitter vacua in the racetrack case since the constraint on the compactified volume size is very much relaxed.
So the model admits solutions with a large adjustable volume.

Interestingly, in this Racetrack K\"ahler uplift model, the stability condition for both the real and imaginary sectors requires that the minima of the potential $V$ always exist for $\Lambda \ge 0$  at large volumes.
Further, the cosmological constant $\Lambda$ is naturally exponentially suppressed as a function of the volume size, and the resultant probability distribution $P(\Lambda)$ for $\Lambda$ gets a sharply peaked behavior toward $\Lambda \rightarrow  0$, which can be highly divergent..
This peaked behavior of $P(\Lambda)$ can be much sharper than that of the previous K\"ahler Uplift model with a single non-perturbative term studied in \cite{Sumitomo:2012vx}. So an exponentially small median for $\Lambda$ is natural.

The racetrack K\"ahler uplift model is similar to the above K\"ahler Uplift model, but with one major addition. The super-potential  $W$ now has two non-perturbative terms for the K\"ahler modulus $T=t + i\tau$ instead of one,
 \begin{equation}
W= W_0 (U_i, S) + W_{\rm NP} = W_0 (U_i, S) + A e^{- a T} + B e^{-b T} 
\end{equation}
where the coefficients $a=2\pi/N_1$ for $SU(N_1)$ gauge symmetry and $b=2\pi/N_2$ for $SU(N_2)$ gauge symmetry. 
In the large volume region and in units where $M_P=1$, the resulting potential may be approximated to
\begin{equation}
 \begin{split}
  &V \simeq \left(-{a^3 A \, W_0\,  \over 2}\right) \lambda (x,y), \\
  & \lambda (x,y) =  - {e^{-x} \over x^2} \cos y - {\beta \over z} {e^{-\beta x} \over x^2} \cos (\beta y) + {\hat{C} \over x^{9/2}},\\
  &x = a t, \quad y=a \tau, \quad   z = A/B, \quad \beta = b/a=N_1/N_2 >1, \quad {\hat{C}} = -{3 a^{3/2} W_0 \, \xi \over 32 \sqrt{2} A}, 
 \end{split}
 \label{approximated potential}
\end{equation}
The extremal conditions $\partial_t V = \partial_{\tau} V = 0$ may be solved.
The typical values of $a$, $\beta$ and $x$ are ${\cal O}(2\pi/16)$, ${\cal O}(1)$ and ${\cal O}(100)$ respectively, and the $e^{-x}$ factor suggests very small $\Lambda$ as well as moduli masses. After randomizing $W_0$, $A$ and $B$, we collect the solutions and find that
the probability distribution $P(\Lambda)$ for small positive $\Lambda$ is approximately given by \cite{Sumitomo:2013vla},
\begin{equation}
 P(\Lambda) \stackrel{\Lambda \rightarrow 0}{\sim} {243 \beta^{1/2} \over 16 (\beta-1)} {1 \over \Lambda^{\beta+1 \over 2 \beta}  (-\ln \Lambda)^{5/2}}.
  \label{asymptotics of PDF}
\end{equation}
So for $\beta \gtrsim 1$, we see that the diverging behavior of $P(\Lambda)$ is very peaked as  $\Lambda \rightarrow 0$.
Since $(\beta +1)/2 \beta < 1$, $P(\Lambda)$ is normalizable, i.e.,$\int P(\Lambda) d \Lambda =1$. For illustration, we have 
\begin{align}
\beta &=1.10 : \quad \quad  \Lambda_{50} = 7.08 \times 10^{-10}, \quad \Lambda_{10} = 3.61 \times 10^{-24}\\
 \beta &=1.04 :   \quad \quad  \Lambda_{50}= 5.47 \times 10^{-19}, \quad \Lambda_{10} = 2.83 \times 10^{-54}
\end{align}
where $\beta=N_1/N_2=26/25=1.04$.
Choosing a gauge group larger than $N_2=25$, $\beta =N_1/N_2=(N_2+1)/N_2$ is closer to unity and the median $\Lambda$ will take values much closer to the the observed value (\ref{L1}). 
We also see that both $t$ and $\tau$ masses are exponentially suppressed. By using the small value of $\Lambda$, we can obtain bounds on both masses \cite{Tye:2016jzi},
\begin{align}
\frac{m^2_t}{\Lambda} &= \frac{\partial_t^2 V}{2K_{T\bar{T}}\Lambda} \leq  \frac{  9\beta x+30(\beta+1)}{a^4 (  2\beta x-5(\beta+1))}, \nonumber \\
\frac{m^2_{\tau}}{\Lambda} &= \frac{\partial_{\tau}^2 V}{2K_{T\bar{T}}\Lambda}\leq \frac{6 x ( 3 \beta x+10(\beta+1))}{a^4 \left(4 \beta  x^2-10 (\beta +1) x+35\right)}.
\end{align}
Solving for $x \sim {\cal O}(100)$, we see that the K\"ahler modulus masses are exponentially small unless one fine-tunes one of the denominating factor to a very small value. Presumably there are heavier scalar bosons that have been integrated out in this simple model. Although their masses are much heavier than $t$ and $\tau$, some of them   can easily have masses that are much smaller than the string or Planck scale.

\section{Discussions}

So far, we have a few looks at the global picture of some corners of the string landscape. As illustrated by the K\"ahler uplift models and the racetrack model discussed, we see hints that, of the meta-stable solutions, most of them have very small $\Lambda$, while each such vacuum has very light bosons.  Here we like to discuss a few issues related to this property.

Since $P(\Lambda)$ is properly normaiized, even with a devergent $P(\Lambda=0)= \infty$, the probability of having a universe with $\Lambda \equiv 0$ is still zero. Furthermore, the flux values are discrete, so the probability distribution $P(a)$ of flux $a$ is non-zero only at discrete values of $a$. That is, $\Lambda$ can be exponentially small, but highly unlikely to be exactly zero.  

The string theory models studied in this paper are admittedly relatively simple. Nonetheless, they incorporate known stringy properties in a consistent fashion so they are non-trivial enough for us to learn about the structure and dynamics of flux compactification in string theory. They clearly illustrate that a statistical preference for a very small physical $\Lambda$ in the cosmic landscape as a solution to the cosmological constant problem is a distinct possibility. This way to solve the cosmological constant problem bypasses the radiative instability problem. Associated with the very small $\Lambda$ are very light moduli masses. So this offers the possibility of having light bosons via statistical preference as well. It is important to point out that this solution or explanation is possible because of the existence of the landscape. Comparing to the earlier works \cite{Bousso:2000xa} where explicit interactions among the moduli and fluxes are not taken into account, we see that the statistical preference for a small $\Lambda$ (and at times some scalar masses) emerges only when dynamics is included. Intuitively, in examining the models studied (albeit a rather limited sample),  more fluxes and moduli coupled together tend to enhance or at least maintain the divergence of $P(\Lambda)$ at $\Lambda=0$. This is encouraging, since higher order corrections and more realistic (and so more complicated) models are very challenging to study. 

What happens when finite temperature and phase transition appear ?
Suppose the Universe starts out at a random point somewhere high up in the landscape, at
zero temperature (for zero temperature, we mean zero thermal temperature, not the Gibbons-Hawking
temperature $H/2\pi = \sqrt{2V/3\pi}/M_P$, which is assumed to be negligible here). It rolls
down and ends up in a local minimum. Because it starts from a random point, this minimum
may be considered to be randomly chosen. If most of the vacua have a small $\Lambda$, it is likely that
this minimum is one of these small $\Lambda$ vacua.
What happens if we turn on a finite temperature $T$ ? We have essentially the same landscape, but is starting from a different point up in the landscape, so the evolution of
the Universe will be different and possibly ending at a different local minimum, also randomly
chosen. As temperature $T \rightarrow 0$, we find that the chosen local vacuum at $T$ probably turns out
to have a small $\Lambda$ at $T=0$, because most vacua at $T=0$ have a small $\Lambda$. If the chosen local
vacuum has a critical temperature $T_c<T$, phase transition happens as $T$ drops below $T_c$. If this
is a second order phase transition, then the Universe will roll away to another local minimum,
which is likely to have a small $\Lambda$ as $T \rightarrow 0$, because most vacua at $T=0$ have a small $\Lambda$. If it is
a first order phase transition, the Universe will stay at this vacuum as $T \rightarrow 0$ (before tunneling).
This vacuum should have a small $\Lambda$, because most vacua at $T=0$ have a small $\Lambda$. In all cases,
we see that the Universe most likely end up in a vacuum with a small $\Lambda$. It is possible that this
same vacuum has a relatively large $\Lambda$ at finite $T$. 

In terms of cosmology, one may wonder why the dark energy is so large, contributing to about $70\%$ of the content of our universe. However, from the fundamental physics point of view, the puzzle is why it is so small, when we assume that the scale of gravity is dictated by the Planck scale $M_P$ which is so much bigger. In our new picture, once we are willing to accept that the smallness of $\Lambda$ has a fundamental explanation like the statistical preference employed here, the question is again reversed. For example, in the viewpoint adopted here, we see that typical moduli mass scales are guided by $\Lambda$, not $M_P$. That is, some of the bosonic/axionic masses are small enough to play the role of fuzzy dark matter.

Once we accept that both $\Lambda$ and $M_P$ have their respective places in the theory (that is, generated by string theory dynamics ,with string scale $M_S$, not via fine-tuning), the presence of some intermediate mass scales such as the Higgs boson mass should not be so surprising. We see that the probability distribution $P(m^2)$ of bosonic mass $m^2$ does not peak at $m^2=0$ in the $\phi^3 / \phi^4$ model. In the string theory models, one envisions scenarios where some bosonic masses have a statistical preference for small values, but such preference is not as strong as that for $\Lambda$. So the Higgs mass $m_H =125$ GeV may fit in in such a scenario, thus evading the usual mass hierarchy problem for the Higgs boson. The scenario also offers the possibility that very light bosons can be present as the dark matter in our universe. In fact, any small number (e.g., the $\theta$ angle, light quark or neutrino masses in the standard electroweak model) in nature may be due to some level of a statistical preference without fine-tuning. Of course, we expect that heavier scalar bosons are generically present in a realistic string model, but they may have been integrated out in the low energy effective theory studied here. One can imagine that some of these "heavy" scalar bosons have masses order-of-magnitude smaller than the string or the Planck scales. 

The string theory models considered so far are necessarily relatively simple, to allow semi-analytic studies. It will be important to consider more realistic versions (for example, the form of the K\"ahler potential and couplings among moduli) to see if such statistical preference for small $\Lambda$ and small bosonic masses are robust. 
In the search for the standard model within string theory, it may be fruitful to narrow the search of the three family standard model only in the region of the landscape where order of magnitude mass scales as well as $\Lambda$ come out in the correct range.

\vspace{0.4cm}

\noindent {\bf Acknowledgments}

\vspace{0.3cm}

I am grateful to Yoske Sumitomo and Sam Wong for their collaborations.
I also thank my HKUST colleagues for discussions in preparing this talk.

\end{document}